\documentclass[12pt,prd,tightenlines,nofootinbib,showpacs,showkeys]{revtex4}
\usepackage{graphics}
\usepackage{rotating}
\usepackage{epsfig}
\usepackage{amsmath}
\usepackage{amsfonts}
\newcommand{\be}{\begin{equation}}
\newcommand{\ee}{\end{equation}}
\newcommand{\bdis}{\begin{displaymath}}
\newcommand{\edis}{\end{displaymath}}
\newcommand{\bes}{\begin{equation*}}
\newcommand{\ees}{\end{equation*}}

\newcommand{\1}{\varepsilon_1}
\newcommand{\2}{\varepsilon_2}
\newcommand{\ep}{\varepsilon_P^\ast}
\newcommand{\eq}{\varepsilon_Q^\ast}
\newcommand{\p}{\!\cdot\!}
\allowdisplaybreaks
\begin{document}

\title{\begin{flushright}{\rm\normalsize SSU-HEP-12/07\\[5mm]}\end{flushright}
Relativistic corrections to double charmonium production \\ in high energy
proton-proton interaction}
\author{\firstname{A.P.} \surname{Martynenko}}
\email{a.p.martynenko@samsu.ru}
\affiliation{Samara State University, Pavlov Street 1, 443011, Samara, Russia}
\affiliation{Samara State Aerospace University named after S.P. Korolyov, Moskovskoye Shosse 34, 443086,
Samara, Russia}
\author{\firstname{A.M.} \surname{Trunin}}
\email{amtrnn@gmail.com}
\affiliation{Samara State Aerospace University named after S.P. Korolyov, Moskovskoye Shosse 34, 443086,
Samara, Russia}

\begin{abstract}
On the basis of pertubative QCD and the relativistic quark model we calculate relativistic
corrections to the process of pair $J/\psi$ production in proton-proton collisions at LHC energy
$\sqrt S=7$~TeV. Relativistic terms in the production amplitude connected with the relative
motion of heavy quarks and the transformation law of the bound state wave functions to the
reference frame of moving $J/\psi$ mesons are taken into account. For the gluon and quark
propagators entering the amplitude we use a truncated expansion in relative quark momenta
up to the second order. Relativistic corrections to the quark bound state wave functions
are considered by means of the Breit-like potential. It turns out that the examined
effects decrease initial nonrelativistic cross section more than two times.
The final result lies below the experimental value measured by LHCb.
\end{abstract}



\maketitle

\section{Introduction}

The production of heavy quarkonium states in different reactions is the subject
of considerable interest during last years. The mechanism of heavy quarkonium
production represents the long-standing problem of quantum
chromodynamics~\cite{likhoded1988,kramer,braaten,brambilla2011}.
Most current theoretical investigations are performed on the basis of
nonrelativistic quantum chromodynamics (NRQCD)~\cite{BBL} and the quark models.
According to these approaches the production of heavy quarkonium
is divided into two stages. On the first stage one or several quark-antiquark
pairs are produced at small distances of order $1/m_Q$. This short-range part
in the production is associated with the basic interaction of free quarks and gluons
and can be evaluated by perturbation theory.
The subsequent nonperturbative transition from the intermediate states
of quarks $QQ...$ and antiquarks $\bar Q\bar Q...$ to physical quarkonium states, on the
second stage, involves long-distance scales of order of quarkonium size $1/(m_Q v)$. The
formation of the quark bound states is parameterized by nonperturbative matrix elements in NRQCD
and calculated by means of the bound state wave functions in the quark models. The finding of
the correspondence between parameters of the quark models and NRQCD opens the way for better
understanding of the quark-gluon dynamics. Both approaches complement each other and can reveal
new aspects of the color dynamics of quarks and gluons.

One of the directions in this investigation is related with the pair production of double
heavy quarkonium. The initial impulse to intensive investigations
was given in this field several years ago by the
measurement of the pair charmonium production cross sections in electron-positron annihilation.
The experimental data obtained at the Belle and BaBar detectors disagreed with the
calculations based on NRQCD. The theoretical results were improved and adjusted in correspondence
with the experiment after the account of one-loop perturbative corrections and relativistic
corrections to the nonrelativistic cross section~\cite{bodwin1,chao1,ebert1}. One of the learned
lessons from this problem
consists in the understanding that only sequential relativistic approach to the heavy quarkonium
production processes can lead to reliable theoretical results. It is necessary to point out that
subsequent observation of numerous charmonium-like states by the Belle and BaBar collaborations with
unclear nature demands further continuation of the investigations in this direction~\cite{pahlova}.
Recently, the first experimental result of the LHCb collaboration on the pair charmonium production
in proton-proton interaction was published~\cite{LHCb}:
\be
\label{eq:lhcb-num}
\sigma^{exp}_{LHCb}=5.1\pm 1.0\pm 1.1~\text{nb},
\ee
where the first uncertainty is statistical and the second systematic.
It agrees with the theoretical estimation of the cross section $\sigma=17\div 22$~nb obtained
in the leading order of QCD
where the process of the gluon fusion is the dominant one~\cite{berezhnoy,baranov,li,ko,qiao}.
These calculations give the total value of the cross section for the pair charmonium production
$\sigma=3\div 5$~nb in the kinematical region of the LHCb experiment (the region of rapidities
\hbox{$2< y< 4.5$}).
The theoretical uncertainty remains sufficiently large. To the present the calculations of
the pair charmonium production in $pp$ interaction were carried out in the leading order of QCD
without inclusion of relativistic corrections. In addition to these permanent theoretical errors
known from the experience of $e^+e^-$ annihilation we have in this task new specific uncertainty
caused by parton distribution functions at small $x$ values because the gluon contribution from the
region of small $x$ is dominant. There appears also another uncertainty related with the double
parton interaction~\cite{baranov1}. In this work we study one aspect of the improvement of the previous
calculations
connected with the account of relativistic corrections. Using the methods of the relativistic
quark model~\cite{ebert1,apm2005,ebert2,ebert3} we perform new calculation of the cross section
$\sigma(pp\to J/\psi J/\psi)$
accounting for relativistic corrections to the production amplitude and the bound state wave
functions of heavy $c$-quarks. So, the aim of this study consists in the relativistic description
of the pair charmonium production at hadron colliders. It is important to note that the interest
to the inclusive reactions $p+p\to 2J/\psi+X$, $pN\to 2J/\psi+X$ is not limited only by the
investigation of the production mechanism. The study of the quarkonium production in the
nuclear matter leads to new data about QCD at high density and temperature~\cite{brodsky1,NPB}.

There exist different mechanisms for the pair charmonium production in $pp$-collisions.
At the collider energies, double quarkonium production occurs through the gluon-gluon interaction channel.
In the color
singlet model (CSM) a pair of quark-antiquark $(c\bar c)$ is created at short distances in color singlet
state. Then it evolves nonperturbatively into an observed meson $J/\psi$. At small transverse momenta
and small invariant masses of the $J/\psi$ pair the color singlet mechanism gives the main
contribution to the cross section. Another possibility is to create a pair $(c\bar c)$ with different
color and angular momentum from that of the final meson. Then the color octet pair evolves to the
color singlet charmonium emitting soft gluons. This color octet mechanism (COM) plays significant role
in the region of high transverse momentum. Among large number of the Feynman diagrams describing the
production of a pair $J/\psi$ it has been separated a class in which the $J/\psi$ pair production
is related with the double gluon fragmentation. In this study we analyze the total set of the production
amplitudes in the color singlet model making primary emphasis upon relativistic effects.

\section{General formalism}

The differential cross section $d\sigma$ for the inclusive double charmonium production in
proton-proton interaction can be presented in the form of the convolution of partonic cross
section $d\sigma(gg\to J/\psi J/\psi)$ with the parton distribution functions (PDF) in the
initial protons~\cite{likhoded1988,kramer}:
\be
\label{eq:cs-plus-x}
d\sigma[p+p\to J/\psi+ J/\psi+X]=\int \! dx_1 dx_2 \, f_{g/p}(x_1,\mu) f_{g/p}(x_2,\mu)
\, d\sigma[gg\to J/\psi J/\psi],
\ee
where $f_{g/p}(x,\mu)$ is partonic distribution function for the gluon in the proton, $x_{1,2}$ are
longitudinal momentum fractions of gluons, $\mu$ is the factorization scale.
Neglecting the proton mass and taking
the c.m. reference frame of initial protons with the beam along the $z$-axis
we can present the gluon on mass-shell momenta $k_{1,2}=x_{1,2}\frac{\sqrt{S}}{2}(1,0,0,\pm 1)$.
$\sqrt{S}$ is the center-of-mass energy in proton-proton collision.

In the quasipotentional approach the invariant transition amplitude for the gluonic subprocess
$g+g\to J/\psi+J/\psi$ can be presented as a convolution of a perturbative production amplitude
of two $c$-quark and $\bar c$-antiquark pairs $\mathcal T(p_1,p_2;q_1,q_2)$ and the quasipotential
wave functions of the final mesons $\Psi_{J/\psi}$~\cite{ebert1}:
\be
\label{eq:m-gen}
{\mathcal M}[gg\to J/\psi J/\psi](k_1,k_2,P,Q)=\int \! \frac{d\mathbf p}{(2\pi)^3}
\int \! \frac{d\mathbf q}{(2\pi)^3} \, \bar\Psi(p,P) \bar\Psi(q,Q) \otimes \mathcal T(p_1,p_2;q_1,q_2),
\ee
where $p_1$ and $p_2$ are four-momenta of $c$-quark and $\bar c$-antiquark in the pair
forming the first $J/\psi$ particle, and $q_2$ and $q_1$ are appropriate momenta for
quark and antiquark in the second meson $J/\psi$. They are defined in terms of total momenta
$P(Q)$ and relative momenta $p(q)$ as follows:
\be
p_{1,2}=\frac12 P \pm p,\quad (pP)=0; \qquad q_{1,2}=\frac12 Q \pm q,\quad (qQ)=0,
\ee
In Eq.~\eqref{eq:m-gen} we integrate over the relative three-momenta of quarks and antiquarks
in the final state.
The systematic account of all terms depending on the relative quark momenta $p$ and $q$
in~\eqref{eq:cs-plus-x} is
important for increasing the accuracy of the calculation.
$p=L_P(0,\mathbf p)$ and $q=L_Q(0,\mathbf q)$
are the relative four-momenta obtained by the Lorentz transformation of four-vectors
$(0,\mathbf p)$ and $(0,\mathbf q)$ to the reference frames moving with the four-momenta $P$~and~$Q$.
The parton-level differential cross section for $g+g\to J/\psi+J/\psi$ is expressed further through
the Mandelstam variables $s$, $t$ and $u$:
\be
\label{eq:stu-def}
s=(k_1+k_2)^2=(P+Q)^2=x_1x_2S,
\ee
\bdis
t=(P-k_1)^2=(Q-k_2)^2=M^2-x_1\sqrt{S}(P_0-|{\bf P}|\cos\phi)=M^2-x_1x_2S+x_2\sqrt{S}(P_0+|{\bf P}|\cos\phi),
\edis
\bes
u=(P-k_2)^2=(Q-k_1)^2=M^2-x_2\sqrt{S}(P_0+|{\bf P}|\cos\phi)=M^2-x_1x_2S+x_1\sqrt{S}(P_0-|{\bf P}|\cos\phi),
\ees
where $M$ is the charmonium mass, $\phi$ is the angle between ${\bf P}$ and the $z$-axis.
The transverse momentum $P_T$ of $J/\psi$ and its energy $P_0$ can be written as
\be
\label{eq:pt0-def}
P_T^2=|{\bf P}|^2\sin^2\!\phi=-t-\frac{(M^2-t)^2}{x_1x_2S},\quad P_0=\frac{x_1x_2\sqrt{S}}{x_1+x_2}+\frac{x_1-x_2}{x_1+x_2}|{\bf P}|\cos\phi.
\ee

\begin{figure}[t]
\center\includegraphics[scale=0.8]{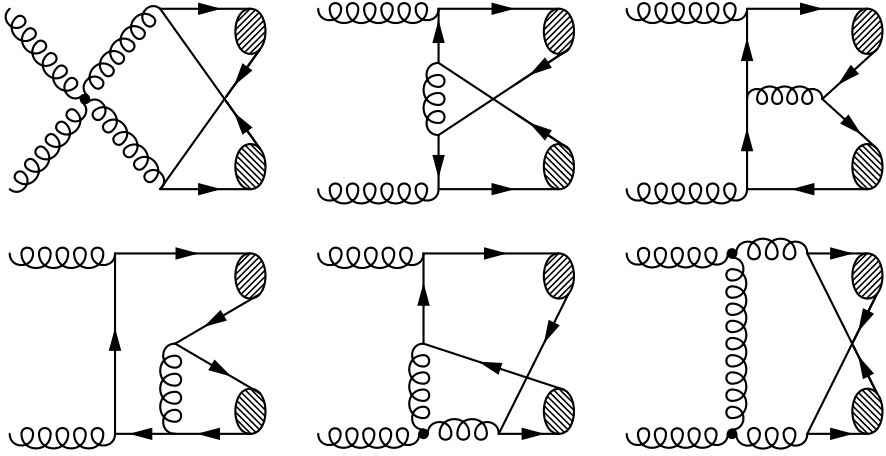}
\caption{The typical LO diagrams contributing to the partonic process $g+g\to J/\psi +J/\psi$.
The others can be obtained by reversing the quark lines or interchanging the initial gluons.}
\label{fig1:diags}
\end{figure}

\begin{figure}[t]
\center\includegraphics[scale=0.8]{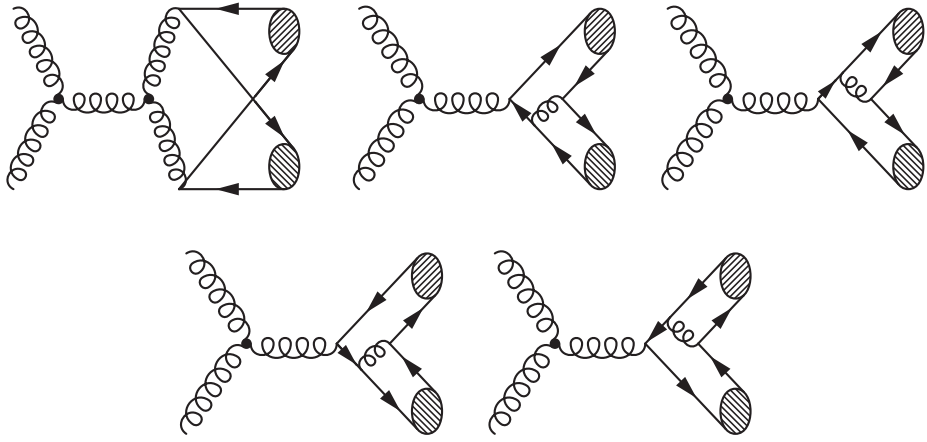}
\caption{The subset of LO diagrams which give zero contribution to $g+g\to J/\psi +J/\psi$
because their color factor is equal zero.}
\label{fig2:diags}
\end{figure}

\begin{figure}[t!]
\center\includegraphics[scale=0.8]{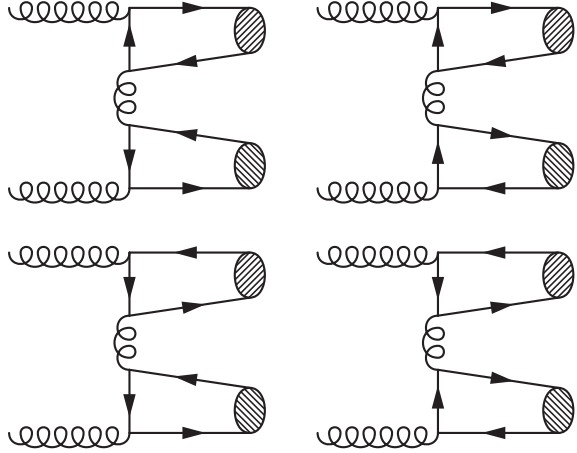}
\caption{The sum of LO diagrams from this subset gives zero contribution to $g+g\to J/\psi +J/\psi$
with the account of relativistic corrections of order $O(v^2)$.}
\label{fig3:diags}
\end{figure}

At leading order of perturbation theory in strong coupling constant $\alpha_s$ there are
31 Feynman diagrams contributing to the amplitude of pair $J/\psi$ production due to gluon fusion.
The typical diagrams from this set are presented in Fig.~\ref{fig1:diags}. For the completeness we
show in Figs.~\ref{fig2:diags}--\ref{fig3:diags} also the Feynman diagrams from two other subsets
containing 5 and 8 Feynman
diagrams which do not contribute to the production amplitude. Any Feynman amplitude shown in
Fig.~\ref{fig2:diags}
has zero contribution because its color factor is equal to zero. The sum of four diagrams from the
subset in Fig.~\ref{fig3:diags} is equal zero with the account of relativistic corrections studied in this work.
So, further we consider only the 31 Feynman amplitudes from Fig.~\ref{fig1:diags}.

Let us consider, for example, the transformation of the first amplitude in Fig.~\ref{fig1:diags} which takes
the form:

\begin{displaymath}
\mathcal T_1^{ab}(p_1,p_2;q_1,q_2)=8\pi^2\alpha_s^2\delta^{ab}\varepsilon_1^\lambda(k_1)
\varepsilon_2^\mu(k_2)\frac{2g_{\lambda\mu}g_{\nu\sigma}-g_{\lambda\sigma}g_{\mu\nu}-g_{\lambda\nu}g_{\mu\sigma}}
{\left(\frac{P}{2}+\frac{Q}{2}+p+q\right)^2\left(\frac{P}{2}+\frac{Q}{2}-p-q\right)^2}\times
\end{displaymath}
\be
\label{eq:t1}
[\bar u(p_1)\gamma^\sigma v(q_1)][\bar u(q_2)\gamma^\nu v(p_2)]
\ee
where $\varepsilon_1(k_1)$ and $\varepsilon_2(k_2)$ are the polarization vectors of initial gluons.
The amplitude~\eqref{eq:m-gen}
should be convoluted with two wave functions of $J/\psi$ mesons taking in the reference frame moving with
four momenta $P$ and $Q$. The transformation law of the bound state wave function from the rest
frame to the moving one with four momentum $P$ was derived in the Bethe-Salpeter approach in Ref.~\cite{brodsky2}
and in the quasipotential method in Ref.~\cite{faustov}. We use the last one and write the
necessary transformation as follows:
\begin{equation}
\Psi_{P}^{\rho\omega}({\bf p})=D_1^{1/2,~\rho\alpha}(R^W_{L_{P}})
D_2^{1/2,~\omega\beta}(R^W_{L_{P}})\Psi_{0}^{\alpha\beta}({\bf p}),
\end{equation}
\begin{displaymath}
\bar\Psi_{P}^{\lambda\sigma}({\bf p})
=\bar\Psi^{\varepsilon\tau}_{0}({\bf p})D_1^{+~1/2,~\varepsilon
\lambda}(R^W_{L_{P}})D_2^{+~1/2,~\tau\sigma}(R^W_{L_{P}}),
\end{displaymath}
where $R^W$ is the Wigner rotation, $L_{P}$ is the Lorentz boost
from the meson rest frame to a moving one, and the rotation matrix
$D^{1/2}(R)$ is defined by
\begin{equation}
{1 \ \ \,0\choose 0 \ \ \,1}D^{1/2}_{1,2}(R^W_{L_{P}})= S^{-1}({\bf
p}_{1,2})S({\bf P})S({\bf p}),
\end{equation}
where the explicit form for the Lorentz transformation matrix of the
four-spinor is
\begin{equation}
S({\bf p})=\sqrt{\frac{\epsilon(p)+m}{2m}}\left(1+\frac{(\boldsymbol\alpha
{\bf p})}{\epsilon(p)+m}\right).
\end{equation}
Omitting a number of transformations which can be performed with~\eqref{eq:t1}
in~\eqref{eq:m-gen}
as in~\cite{ebert1,ebert2} we write the contribution to the production amplitude in the form:
\be
\label{eq:m1-dec}
{\mathcal M_1}^{ab}(k_1,k_2,P,Q)=
2 M\delta^{ab}\pi^2\alpha_s^2\int\frac{d\mathbf p}{(2\pi)^3}
\frac{\bar\Psi_0^{J/\psi}({\bf p})}{[\frac{\epsilon(p)}{m}\frac{\epsilon(p)+m}{2m}]}
\int\frac{d\mathbf q}{(2\pi)^3}\frac{\bar\Psi_0^{J/\psi}({\bf q})}{[\frac{\epsilon(q)}{m}\frac{\epsilon(q)+m}{2m}]}
\times
\ee
\begin{displaymath}
\mathrm{Tr}\Biggl\{\left[\frac{\hat v_1-1}{2}+\hat v_1\frac{{\bf p}^2}{2m(\epsilon(p)+m)}-
\frac{\hat p}{2m}\right]\hat\varepsilon_P^\ast(\hat v_1+1)\left[\frac{\hat v_1+1}{2}+
\hat v_1\frac{{\bf p}^2}{2m(\epsilon(p)+m)}+
\frac{\hat p}{2m}\right]\gamma^\sigma\times
\end{displaymath}
\begin{displaymath}
\left[\frac{\hat v_2-1}{2}+\hat v_2\frac{{\bf q}^2}{2m(\epsilon(q)+m)}+
\frac{\hat q}{2m}\right]\hat\varepsilon_Q^\ast(\hat v_2+1)\left[\frac{\hat v_2+1}{2}+
\hat v_2\frac{{\bf q}^2}{2m(\epsilon(q)+m)}-
\frac{\hat q}{2m}\right]\gamma^\nu\Biggr\}\times
\end{displaymath}
\begin{displaymath}
\times\varepsilon_1^\lambda(k_1)
\varepsilon_2^\mu(k_2)\frac{2g_{\lambda\mu}g_{\nu\sigma}-g_{\lambda\sigma}g_{\mu\nu}-g_{\lambda\nu}g_{\mu\sigma}}
{\left(\frac{P}{2}+\frac{Q}{2}+p+q\right)^2\left(\frac{P}{2}+\frac{Q}{2}-p-q\right)^2},
\end{displaymath}
where
$v_1=P/M$, $v_2=Q/M$; $\varepsilon_{P,Q}$ are polarization vectors
of final $J/\psi$ mesons with $\varepsilon_P\cdot P=0$ and $\varepsilon_Q\cdot Q=0$;
$\epsilon(p)=\sqrt{m^2+\mathbf p^2}$, $M=2m$
is the mass of $J/\psi$ meson.
The hat symbol means contraction of the four-vector with the Dirac gamma-matrices.
The spin projectors
$v(0)\bar u(0)=\hat\varepsilon^\ast(1+\gamma_0)/(2\sqrt 2)$
corresponding to $J/\psi$ mesons are introduced as well as projectors $\delta_{ij}/\sqrt 3$
onto a color singlet states. We explicitly extracted in~\eqref{eq:m1-dec} the normalization factors
$\sqrt{2M}$ of quasipotential bound state wave functions.

The same transformations can be carried out with all 31 Feynman amplitudes in Fig.~\ref{fig1:diags}
In view of large volume of the calculations we have used the package FeynArts~\cite{feynarts}
for the system Mathematica and Form~\cite{form}. To make the entry of final
amplitude more compact, we introduce a number of vertex functions $\Gamma_i$.
Then we can present the total amplitude~\eqref{eq:m-gen} in the form:
\begin{equation}
\label{eq:m-part}
{\mathcal M}[gg\to J/\psi J/\psi](k_1,k_2,P,Q)=
\frac19 M\,\pi^2\alpha_s^2\int\frac{d\mathbf p}{(2\pi)^3}\int\frac{d\mathbf q}{(2\pi)^3}
\, \mathrm{Tr} \, \mathfrak M,
\ee
\bdis
\mathfrak M=
\mathcal D_{1} \gamma_\beta \bar\Psi_{q,Q} \Gamma_{1}^\beta \bar\Psi_{p,P}\,
  \hat{\varepsilon}_2 \frac{m - \hat{k}_2 + \hat{p}_1}{(k_2-p_1)^2-m^2} +
\mathcal D_{2} \gamma_\beta \bar\Psi_{q,Q} \Gamma_{2}^\beta \bar\Psi_{p,P}\,
  \hat{\varepsilon}_1 \frac{m - \hat{k}_1 + \hat{p}_1}{(k_1-p_1)^2-m^2} +
\edis
\bdis
\mathcal D_{3} \bar\Psi_{q,Q} \Gamma_{3}^\beta \bar\Psi_{p,P}\, \gamma_\beta +
\mathcal D_{4} \bar\Psi_{p,P} \Gamma_{4}^\beta \bar\Psi_{q,Q}\, \gamma_\beta +
 \mathcal D_{1} \bar\Psi_{q,Q} \Gamma_{5}^\beta \bar\Psi_{p,P} \,
  \gamma_\beta \frac{m + \hat{k}_2 - \hat{q}_1}{(k_2-q_1)^2-m^2} \hat{\varepsilon}_2 +
\edis
\bdis
 \mathcal D_{2} \bar\Psi_{q,Q} \Gamma_{6}^\beta \bar\Psi_{p,P} \,
  \gamma_\beta \frac{m + \hat{k}_1 - \hat{q}_1}{(k_1-q_1)^2-m^2} \hat{\varepsilon}_1,
\edis
where inverse denominators  of gluon propagators are defined as
$\mathcal D_{1,2}^{-1}=(k_2-p_{1,2}-q_{1,2})^2$ and  $\mathcal D_{3,4}^{-1}=(p_{1,2}+q_{1,2})^2$.
Pertubative amplitude $\mathcal T(p_1,p_2;q_1,q_2)$ in Eq.~\eqref{eq:m-gen} describes
production of two $c$-quark and $\bar c$-antiquark pairs with the momenta $p_{1,2}$ and
$q_{2,1}$ respectively. The formation of observable bound states from quark--antiquark
pairs is determined in the quark model by the quasipotential wave functions $\Psi_{J/\psi}(p,P)$ and
$\Psi_{J/\psi}(q,Q)$. These wave functions are calculated initially
in the meson rest frame and then transformed to the reference frames moving with the four-momenta
$P$ and $Q$. As a result we obtain in~\eqref{eq:m-part}
the following expressions for relativistic bound state
wave functions which determine the transition of heavy quarks to the bound state:
\begin{eqnarray}
\notag
\bar\Psi_{p,P}=\frac{\bar\Psi_0^{J/\psi}(\mathbf p)}{\bigl[\frac{\epsilon(p)}{m}\frac{\epsilon(p)+m}{2m}\bigr]}
\left[
	\frac{\hat v_1-1}{2}+\hat v_1\frac{\mathbf p^2}{2m(\epsilon(p)+m)}-\frac{\hat p}{2m}
\right] \times \\
\label{eq:psi-p}
 \hat\varepsilon^\ast_P(P,S_z) \, (1+\hat v_1) \!
\left[
	\frac{\hat v_1+1}{2}+\hat v_1\frac{\mathbf p^2}{2m(\epsilon(p)+m)}+\frac{\hat p}{2m}
\right],
\end{eqnarray}
\begin{eqnarray}
\notag
\bar\Psi_{q,Q}=\frac{\bar\Psi_0^{J/\psi}(\mathbf q)}{\bigl[\frac{\epsilon(q)}{m}\frac{\epsilon(q)+m}{2m}\bigr]}
\left[
	\frac{\hat v_2-1}{2}+\hat v_2\frac{\mathbf q^2}{2m(\epsilon(q)+m)}+\frac{\hat q}{2m}
\right] \times \\
\label{eq:psi-q}
 \hat\varepsilon^\ast_Q(Q,S_z) \, (1+\hat v_2) \!
\left[
	\frac{\hat v_2+1}{2}+\hat v_2\frac{\mathbf q^2}{2m(\epsilon(q)+m)}-\frac{\hat q}{2m}
\right].
\end{eqnarray}

Leading order vertex functions in~\eqref{eq:m-part} are calculated in the Feynman gauge and can
be presented as follows:
\bdis
\Gamma_{1}^\beta=\hat{\varepsilon}_1 \frac{m-\hat{k}_1+\hat{q}_2}{(k_1-q_2)^2-m^2} \gamma^\beta-8\, \gamma^\beta \frac{m+\hat{k}_1-\hat{p}_2}{(k_1-p_2)^2-m^2} \hat{\varepsilon}_1,
\edis
\bdis
\Gamma_{2}^\beta=\hat{\varepsilon}_2 \frac{m-\hat{k}_2+\hat{q}_2}{(k_2-q_2)^2-m^2} \gamma^\beta-8\, \gamma^\beta \frac{m+\hat{k}_2-\hat{p}_2}{(k_2-p_2)^2-m^2} \hat{\varepsilon}_2 ,
\edis
\be
\label{eq:vert}
\Gamma_{3}^\beta=\hat{\varepsilon}_1 \frac{m-\hat{k}_1+\hat{q}_2}{(k_1-q_2)^2-m^2}\Bigl[\gamma^\beta \frac{m+\hat{k}_2-\hat{p}_2}{(k_2-p_2)^2-m^2} \hat{\varepsilon}_2-8 \, \hat{\varepsilon}_2
\frac{m-\hat{p}_1-\hat{p}_2-\hat{q}_1}{(p_1+p_2+q_1)^2-m^2} \gamma^\beta\Bigr]+
\ee
\bdis
\hat{\varepsilon}_2 \frac{m-\hat{k}_2+\hat{q}_2}{(k_2-q_2)^2-m^2} \Bigl[\gamma^\beta \frac{m+\hat{k}_1-\hat{p}_2}{(k_1-p_2)^2-m^2} \hat{\varepsilon}_1-8 \, \hat{\varepsilon}_1 \frac{m-\hat{p}_1-\hat{p}_2-\hat{q}_1}{(p_1+p_2+q_1)^2-m^2} \gamma^\beta\Bigr]-
\edis
\bdis
8 \, \gamma^\beta \frac{m+\hat{p}_1+\hat{q}_1+\hat{q}_2}{(p_1+q_1+q_2)^2-m^2} \Bigl[\hat{\varepsilon}_2 \frac{m+\hat{k}_1-\hat{p}_2}{(k_1-p_2)^2-m^2} \hat{\varepsilon}_1+ \hat{\varepsilon}_1 \frac{m+\hat{k}_2-\hat{p}_2}{(k_2-p_2)^2-m^2} \hat{\varepsilon}_2\Bigr]+
\edis
\bdis
18\, \gamma_\alpha \Bigl[\mathcal D_{1} \frac{m-\hat{k}_1+\hat{q}_2}{(k_1-q_2)^2-m^2} \varepsilon_1^\alpha\gamma_\mu\mathfrak{E}_2^{\beta\mu}(p_1+q_1)-\mathcal D_{1} \frac{m+\hat{k}_1-\hat{p}_2}{(k_1-p_2)^2-m^2} \hat{\varepsilon}_1 \mathfrak{E}_2^{\beta\alpha}(p_1+q_1)+
\edis
\bdis
\mathcal D_{2} \frac{m-\hat{k}_2+\hat{q}_2}{(k_2-q_2)^2-m^2} \varepsilon_2^\alpha\gamma_\mu\mathfrak{E}_1^{\beta\mu}(p_1+q_1)- \mathcal D_{2} \frac{m+\hat{k}_2-\hat{p}_2}{(k_2-p_2)^2-m^2} \hat{\varepsilon}_2 \mathfrak{E}_1^{\beta\alpha}(p_1+q_1)\Bigr] ,
\edis
\bdis
\Gamma_{4}^\beta=\hat{\varepsilon}_1 \frac{m-\hat{k}_1+\hat{p}_1}{(k_1-p_1)^2-m^2} \Bigl[\gamma^\beta \frac{m+\hat{k}_2-\hat{q}_1}{(k_2-q_1)^2-m^2} \hat{\varepsilon}_2-8 \, \hat{\varepsilon}_2 \frac{m-\hat{p}_2-\hat{q}_1-\hat{q}_2}{(p_2+q_1+q_2)^2-m^2} \gamma^\beta\Bigr]+
\edis
\bdis
\hat{\varepsilon}_2 \frac{m-\hat{k}_2+\hat{p}_1}{(k_2-p_1)^2-m^2} \Bigl[\gamma^\beta \frac{m+\hat{k}_1-\hat{q}_1}{(k_1-q_1)^2-m^2} \hat{\varepsilon}_1-8 \, \hat{\varepsilon}_1 \frac{m-\hat{p}_2-\hat{q}_1-\hat{q}_2}{(p_2+q_1+q_2)^2-m^2} \gamma^\beta\Bigr]-
\edis
\bdis
8 \,\gamma^\beta \frac{m+\hat{p}_1+\hat{p}_2+\hat{q}_2}{(p_1+p_2+q_2)^2-m^2} \Bigl[\hat{\varepsilon}_2 \frac{m+\hat{k}_1-\hat{q}_1}{(k_1-q_1)^2-m^2} \hat{\varepsilon}_1+
\hat{\varepsilon}_1 \frac{m+\hat{k}_2-\hat{q}_1}{(k_2-q_1)^2-m^2} \hat{\varepsilon}_2\Bigr]+
\edis
\bdis
18\, \gamma_\alpha \Bigl[\mathcal D_{2} \frac{m-\hat{k}_1+\hat{p}_1}{(k_1-p_1)^2-m^2} \varepsilon_1^\alpha\gamma_\mu\mathfrak E_2^{\beta\mu}(p_2+q_2)-
\mathcal D_{2}  \frac{m+\hat{k}_1-\hat{q}_1}{(k_1-q_1)^2-m^2} \hat{\varepsilon}_1\mathfrak E_2^{\beta\alpha}(p_2+q_2)+
\edis
\bdis
\mathcal D_{1} \frac{m-\hat{k}_2+\hat{p}_1}{(k_2-p_1)^2-m^2} \varepsilon_2^\alpha\gamma_\mu\mathfrak E_1^{\beta\mu}(p_2+q_2)-
\mathcal D_{1} \frac{m+\hat{k}_2-\hat{q}_1}{(k_2-q_1)^2-m^2} \hat{\varepsilon}_2\mathfrak E_1^{\beta\alpha}(p_2+q_2)\Bigr]+
\edis
\bdis
18 \, \mathcal D_{3} \gamma_\nu \bigl[2\,\varepsilon_1 \varepsilon_2\, g^{\nu\beta}-\varepsilon_1^\nu \varepsilon_2^\beta-\varepsilon_1^\beta \varepsilon_2^\nu+
\mathcal D_{2} \mathfrak{F}^{\nu\beta}(p_1+q_1,p_2+q_2) +\mathcal D_{1}\mathfrak{F}^{\beta\nu}(p_2+q_2,p_1+q_1)\bigr],
\edis
\bdis
\Gamma_{5}^\beta=\gamma^\beta \frac{m+\hat{k}_1-\hat{p}_2}{(k_1-p_2)^2-m^2} \hat{\varepsilon}_1-8 \, \hat{\varepsilon}_1 \frac{m-\hat{k}_1+\hat{q}_2}{(k_1-q_2)^2-m^2} \gamma^\beta ,
\edis
\bdis
\Gamma_{6}^\beta=\gamma^\beta \frac{m+\hat{k}_2-\hat{p}_2}{(k_2-p_2)^2-m^2} \hat{\varepsilon}_2-8 \, \hat{\varepsilon}_2 \frac{m-\hat{k}_2+\hat{q}_2}{(k_2-q_2)^2-m^2} \gamma^\beta ,
\edis
where we introduce the following tensors:
\be
\begin{gathered}
\mathfrak{E}^{\alpha\beta}_{1,2}(x)=\frac12\bigl(2\,x\varepsilon_{1,2} \, g^{\alpha\beta}-(k_{1,2}^\beta+x^\beta)\varepsilon_{1,2}^\alpha+(2k_{1,2}^\alpha-x^\alpha)\varepsilon_{1,2}^\beta\bigr),\\
\mathfrak{F}^{\alpha\beta}(x,y)=4 (x\varepsilon_1) (y\varepsilon_2) g^{\alpha\beta}+(k_1+x)(k_2+y)\varepsilon_1^\alpha \varepsilon_2^\beta+\varepsilon_1\varepsilon_2(2k_1^\alpha-x^\alpha)(2k_2^\beta-y^\beta)+ \\
2\,x(\varepsilon_1\varepsilon_2^\alpha-\varepsilon_2\varepsilon_1^\alpha)(2k_2^\beta-y^\beta)-
2\,y(\varepsilon_1\varepsilon_2^\beta-\varepsilon_2\varepsilon_1^\beta)(2k_1^\alpha-x^\alpha)- \\
x\varepsilon_1(x^\alpha+4y^\alpha)\varepsilon_2^\beta-y\varepsilon_2(4x^\beta+y^\beta)\varepsilon_1^\alpha.
\end{gathered}
\ee

Our expressions for the amplitude~\eqref{eq:m-part} and vertex
functions~\eqref{eq:vert} contain relative momenta $p$ and $q$ in exact form. In order to take
into account relativistic corrections of the second order in $p$ and $q$ we expand all
inverse denominators of the quark and gluon propagators. Such expansions look as follows:
\be
\begin{gathered}
\label{eq:ex-props}
\frac{1}{(p_1+q_1)^2}=\frac{4}{s}-\frac{16}{s^2} \left[ (p+q)^2+pQ+qP \right] +\cdots, \\
\frac{1}{(k_2-q_2)^2-m^2}=\frac{2}{t-M^2}-\frac{4}{\left(t-M^2\right)^2} \left[ q^2+2\,q k_2\right] +\cdots,
\end{gathered}
\ee
where the Mandelstam variables for the gluonic subprocess $s$ and $t$ are defined in~\eqref{eq:stu-def}.
There are 16 quark and gluon propagators in the amplitude~\eqref{eq:m-part} which have
to be expanded in the same way as in~\eqref{eq:ex-props}. All denominators of
these propagators in nonrelativistic limit take one of the following forms: $(t-M^2)/2$,
$(M^2-s-t)/2$, $\pm s/4$ or $s/2$. Then, the inequalities
\be
4M^2\le s,\quad
\left| t+\frac{s}{2}-M^2 \right|\le\frac{s}{2}\sqrt{1-\frac{4M^2}{s}}
\ee
mean that in the case of the most unfavorable values of the variables $x_{1,2}$ and $t$ we
can roughly estimate expansion parameters in~\eqref{eq:ex-props}
as $2p^2/M^2$ and $2q^2/M^2$.
Preserving in the expanded amplitude terms up to the second order in the relative momenta $p$ and $q$,
we can perform angular integration using the following relations for $\mathcal S$-wave charmonium:
\bdis
\int\!\frac{\Psi^{\mathcal S}_0(\mathbf p)}{\bigl[\frac{\epsilon(p)}{m}\frac{\epsilon(p)+m}{2m}\bigr]}\frac{d\mathbf p}{(2\pi)^3}=\frac{1}{\sqrt{2}\,\pi}\int\limits_0^\infty\!\frac{p^2R_\mathcal S(p)}{\bigl[\frac{\epsilon(p)}{m}\frac{\epsilon(p)+m}{2m}\bigr]}dp,
\edis
\be
\int\! p_\mu p_\nu \, \frac{\Psi^{\mathcal S}_0(\mathbf p)}{\bigl[\frac{\epsilon(p)}{m}\frac{\epsilon(p)+m}{2m}\bigr]}\frac{d\mathbf p}{(2\pi)^3}=-\frac{1}{3\sqrt2\,\pi}(g_{\mu\nu}-{v_1}_\mu{v_1}_\nu)
\int\limits_0^\infty\!\frac{p^4R_\mathcal S(p)}{\bigl[\frac{\epsilon(p)}{m}\frac{\epsilon(p)+m}{2m}\bigr]}dp,
\ee
where $R_\mathcal S(p)$ is the radial charmonium wave function.

To illustrate the described transformations we present here the result of the calculation of the first amplitude
in Fig.~\ref{fig1:diags}:
\bes
\begin{aligned}
&
\mathcal M_1^{ab}=\frac{32\alpha_s^2\delta^{ab}}{9\,m\,s^4}\int\frac{m+\epsilon(p)}{2\epsilon(p)}R(p) p^2 \!\! \int\frac{m+\epsilon(q)}{2\epsilon(q)}R(q) q^2 \biggl\{3s^2\bigl[\1\p\2 (s\,\ep\p\eq-2\,\ep\p Q \, \eq\p P)- \\&
2\,\ep\p\eq(\1\p P \, \2\p Q + \1\p Q \, \2\p P)+ 2\,\ep\p Q (\1\p P \, \2\p\eq + \1\p\eq \, \2\p P)- \1\p\ep \times \\&
(s\,\2\p\eq - 2\,\2\p Q \, \eq\p P) - \2\p\ep(s\,\1\p\eq - 2\,\1\p Q \, \eq\p P)\bigr]\Bigl(3(1-c_p-c_q-c_p^2-c_q^2)+c_pc_q\times\\&
(67+3c_p+3c_q)+3c_p^2c_q^2\Bigr)-64 m^2 s\,\bigl[\ep\p Q (\1\p P\, \2\p\eq + \1\p\eq \, \2\p P) + \eq\p P (\1\p Q \, \2\p\ep+ \\&
\1\p\ep \, \2\p Q)\bigr] \Bigl(3 (c_p + c_q) + c_p c_q (194 - 3 c_p - 3 c_q)\Bigr) + 16 m^2 s^2\,\bigl[\1\p\ep \, \2\p\eq + \1\p\eq \, \2\p\ep\bigr] \times
\end{aligned}
\ees
\be
\begin{aligned}
\label{eq:m1}
&
\Bigl(9 (c_p + c_q) + c_p c_q (380 - 9 c_p - 9 c_q)\Bigr) + 192 m^2 s\,\ep\p \eq \bigl[\1\p P \, \2\p Q + \1\p Q \, \2\p P\bigr] \Bigl(c_p + c_q + \\&
c_p c_q (62 - c_p - c_q)\Bigr)+16 m^2 s\,\1\p \2\,\Bigl(\ep\p \eq \Bigl[32 m^2\,\bigl(3 (c_p + c_q) + c_p c_q (329 - 3 c_p - 3 c_q)\bigr) - 3s\times \\&
\bigl(3 - 2 c_p - 2 c_q - 3 c_p^2 - 3 c_q^2\bigr) + s\,c_p c_q (613 + 6 c_p + 6 c_q) + 9 s\,c_p^2 c_q^2\Bigr] + 4 \,\ep\p Q \, \eq\p P \bigl[3 (c_p + c_q) + \\&
c_p c_q(202 - 3 c_p - 3 c_q)\bigr]\!\Bigr)+512 m^2 c_p c_q \biggl(2\,\ep\p Q \, \eq\p P (\1\p Q \, \2\p P + \1\p P \, \2\p Q) -  2 m^2\,\ep\p \eq \times \\&
\Bigl[1064 m^2\,\1\p \2 + \1\p P (125\, \2\p Q - 8\, \2\p P) + \1\p Q  (125\, \2\p P - 8\, \2\p Q)\Bigr] - \\&
m^2\Bigl[266\, \1\p \2\,\ep\p Q \, \eq\p P + \1\p\ep \bigl(131 s\,\2\p\eq + 2\,\eq\p P (4\, \2\p P - 129\, \2\p Q)\bigr) + \1\p\eq \times \\&
\bigl(131 s\,\2\p\ep+ 2\,\ep\p Q(4\, \2\p Q - 129\, \2\p P)\bigr) + 2\,\1\p P (4\, \2\p\ep\,\eq\p P - 129\, \2\p\eq\,\ep\p Q) + \\&
2\,\1\p Q (4\, \2\p\eq\,\ep\p Q - 129\, \2\p\ep\,\eq\p P)\Bigl]\biggr)\!\biggr\} dp\,dq,
\end{aligned}
\ee
where we introduce the relativistic parameter $c_p=\frac{m-\epsilon(p)}{m+\epsilon(p)}$.
Extracting relativistic factors $(\epsilon +m)/2\epsilon$ in the integrals over both relative
momenta ${\bf p}$ and ${\bf q}$ we observe that the amplitude $\mathcal M_1^{ab}$ is a power-like expansion in
relativistic parameters $c_p$ and $c_q$.
Due to the presence of four different polarization vectors, which correspond to incoming gluons and
outcoming $J/\psi$ particles, the result~\eqref{eq:m1} appears to be sufficiently lengthy. We have also
obtained analogous expressions for remained 30 diagrams, but due to the bulkiness of the total amplitude
they are not presented here.

To calculate the cross section we have to sum the squared modulus of the amplitude upon all polarizations
using the following relations for final $J/\psi$ mesons and initial gluons correspondingly:
\be
\sum_{\lambda}\varepsilon_{P}^\mu \, {\varepsilon_{P}^\ast}^\nu = v_1^\mu v_1^\nu-g^{\mu\nu},\quad
\sum_{\lambda}\varepsilon_{Q}^\mu \, {\varepsilon_{Q}^\ast}^\nu = v_2^\mu v_2^\nu-g^{\mu\nu},\quad
\sum_{\lambda}\varepsilon_{1,2}^\mu \, \varepsilon_{1,2}^{\ast\;\nu}=\frac{
k_1^\mu k_2^\nu+k_1^\nu k_2^\mu}{k_1\cdot k_2}-g^{\mu\nu}.
\ee

We find it useful to present the differential cross section for double charmonium production in
the proton-proton interaction in the following form:
\be
\label{eq:cs}
\frac{d\sigma}{dt}[gg\to J/\psi J/\psi](t,s)=\frac{\pi\:\! m^2\alpha_s^4}{2304\,s^2}\,|\tilde R(0)|^4\sum_{i=0}^3\omega_iF^{(i)}(t,s),
\ee
where the function $F^{(0)}$ describes the LO contribution. It coincides with the
nonrelativistic analytical
expression for the cross section obtained for the studied process in~\cite{berezhnoy,li,qiao,qiao2002}.
The functions $F^{(i)}$ ($i=1,2,3$) describe relativistic corrections.
Explicit expressions for all functions $F^{(i)}$ entering the cross section~\eqref{eq:cs}
are written in Appendix~\ref{app:fis}. A number of specific parameters $\omega_i$
appeared in~\eqref{eq:cs} are defined as
\be
\label{eq:omega-defs}
\omega_0=1,\quad\omega_1=\frac{I_1}{I_0},\quad\omega_2=\frac{I_2}{I_0},\quad\omega_3=\omega_1^2.
\ee
They comprise the nonperturbative parameters in the relativistic quark model which
determine the transition of quarks and antiquarks into the bound states.
The parameter $\tilde R(0)$, which represents the relativistic generalization of radial wave function
at the origin, is defined by the formula:
\be
\tilde R(0)=\sqrt{\frac2\pi}\int\limits_0^\infty\! \frac{m+\epsilon(p)}{2\epsilon(p)}R(p)p^2 dp.
\ee
The parameters $\omega_i$ are determined by integrals containing the bound state wave function
in the following form:
\be
\label{eq:idefs}
I_0=\int\limits_0^\infty \frac{m+\epsilon(p)}{2\epsilon(p)}R(p) p^2 dp,\quad
I_{1,2}=\int\limits_0^m\frac{m+\epsilon(p)}{2\epsilon(p)}\left(\frac{m-\epsilon(p)}
{m+\epsilon(p)}\right)\!\negthickspace{\vphantom{\biggl|}}^{1,2} \! R(p) p^2 dp.
\ee

Our basic relations for the cross section~\eqref{eq:cs} evidently show that there exists another
source of relativistic corrections connected with the charmonium wave functions. For their
calculation with the desired accuracy we suppose that
the dynamics of a $c\bar c$-pair is determined by the QCD generalization of the standard Breit
Hamiltonian~\cite{gupta}, which in the c.m. reference frame can be written as
\be
\label{eq:ham0}
H=H_0+\Delta U_1+\Delta U_2+\Delta U_3,\quad
H_0=2\sqrt{{\bf p}^2+m^2}-2m-\frac{C_F\tilde\alpha_s}{r}+Ar+B,
\ee
\be
\label{eq:delts}
\Delta U_1(r)=-\frac{C_F\alpha_s^2}{4\pi r}\left[2\beta_0\ln(\mu r)+a_1+2\gamma_E\beta_0 \right],\quad
a_1=\frac{31}{3}-\frac{10}{9}n_f,\quad
\beta_0=11-\frac{2}{3}n_f,
\ee
\begin{displaymath}
\Delta U_2(r)=-\frac{C_F\alpha_s}{2m^2r}\left[{\bf p}^2+\frac{{\bf r}({\bf r}{\bf p}){\bf p}}{r^2}\right]+
\frac{\pi C_F\alpha_s}{m^2}\delta({\bf r})+\frac{3C_F\alpha_s}{2m^2r^3}({\bf S}{\bf L})-
\end{displaymath}
\begin{displaymath}
\frac{C_F\alpha_s}{2m^2}\left[\frac{{\bf S}^2}{r^3}-3\frac{({\bf S}{\bf r})^2}{r^5}-\frac{4\pi}{3}(2{\bf S}^2-3)\delta({\bf r})\right]-\frac{C_AC_F\alpha_s^2}{2mr^2},
\end{displaymath}
\begin{displaymath}
\Delta U_3(r)=f_V\Bigl[\frac{A}{2m^2r}\Bigl(1+\frac{8}{3}{\bf S}_1 {\bf S}_2\Bigr)+
\frac{3A}{2m^2r}{\bf L} {\bf S}+\frac{A}{3m^2r}\Bigl(\frac{3}{r^2}({\bf S}_1 {\bf r}) ({\bf S}_2 {\bf r})-
{\bf S}_1 {\bf S}_2\Bigr)\Bigr]-(1-f_V)\frac{A}{2m^2r}{\bf L} {\bf S},
\end{displaymath}
where ${\bf L}=[{\bf r}\times{\bf p}]$, ${\bf S}={\bf S}_1+{\bf S}_2$, $n_f$ is a number of flavors,
$C_A=3$ and $C_F=4/3$ are the color factors of the SU(3) color group, $\gamma_E$ is the Euler constant.
The parameter $f_V$ of vector-exchange confining potential was set to be $f_V=0.7$. The mass of heavy
$c$-quark  in our model is equal to $m=1.55$~GeV. For the dependence of the QCD coupling constant
$\tilde\alpha_s(\mu)$ on the renormalization point $\mu$ in the pure Coulomb term in~\eqref{eq:ham0}
we use the three-loop result~\cite{chetyrkin}
\be
\begin{gathered}
\tilde\alpha_s(\mu)=\frac{4\pi}{\beta_0\mathcal L}-\frac{4\pi b_1\mathcal L}{(\beta_0\mathcal L)^2}+
\frac{4\pi}{(\beta_0\mathcal L)^3}\left[b_1^2(\ln^2\mathcal L-\ln\mathcal L-1)+b_2\right],
\quad \mathcal L=\ln(\mu^2/\Lambda^2),\\
b_1=\frac{64}{9}, \quad b_2=\frac{3863}{54},
\end{gathered}
\ee
whereas in all other terms of the Hamiltonian~\eqref{eq:delts} we use the one-loop approximation
for the coupling constant $\alpha_s$.
The typical momentum transfer scale in a quarkonium is of order of the quark mass, so we choose the
renormalization scale $\mu=m=1.55$ GeV and $\Lambda=0.168$~GeV, which gives $\alpha_s=0.314$ for
the charmonium states. The parameters of the linear potential $A=0.18$ GeV$^2$ and $B=-0.16$ GeV
have the usual values of quark models. Starting with the Hamiltonian~\eqref{eq:ham0} we construct
the effective potential model based on the Schr\"odinger equation and find its numerical solutions
for $J/\psi$ meson. Additional details of this model are contained in Appendix C of \cite{ebert2}.
Note that the obtained charmonium wave function is strongly decreasing in the region of relativistic
momenta $p\gtrsim m$. Our numerical evaluation gives the following values of $\mathcal
S$-wave charmonium masses: $M_{J/\psi}^{th}=3.072$~GeV and $M_{\eta_c}^{th}=2.988$~GeV, which
lie within the reasonable accuracy in the comparison with their experimentally measured
results~\cite{PDG-2010} $M_{J/\psi}^{exp}=3.097$~GeV and $M_{\eta_c}^{exp}=2.980$~GeV.

\section{Numerical results and discussion}

In this work we investigate the role of relativistic effects in the production
of a pair of charmonium mesons in proton-proton interaction in the relativistic
quark model. We have studied only the order $\alpha_s^4$ parton process of gluon-gluon fusion
in the color singlet model.
At the calculation of the production amplitude~\eqref{eq:m-part}
we keep relativistic corrections of two types. The first type is
determined by several functions depending on the relative quark
momenta  ${\bf p}$ and ${\bf q}$ arising from the gluon propagators,
the quark propagators, and relativistic meson wave functions. The
second type of corrections originates from the perturbative and nonperturbative
treatment of the quark-antiquark interaction operator~\eqref{eq:ham0} which leads to
the essential modification of nonrelativistic wave functions.

For the calculation of relativistic corrections in the bound state wave functions
$\Psi_0({\bf p})$ we take the Breit potential~\eqref{eq:delts} and
construct the effective potential model as in~\cite{ebert2,lucha2} by means of
the rationalization of the kinetic energy operator.
Using the program of numerical solution of the Schr\"odinger equation~\cite{lucha1}
we obtain the following values of all
relativistic parameters entering the cross section~\eqref{eq:cs}:
$\tilde R(0)=0.57$~GeV$\strut^\frac32$, $\omega_1=-0.051$ and $\omega_2=0.0047$.

As it is evident from Eq.~\eqref{eq:idefs}, our definition of integral parameters $I_{1,2}$
describing relativistic contributions from the production amplitude contains the cutoff at
relativistic momentum of order $m$. In spite of the convergence of integrals $I_{1,2}$,
our relativistic model can not provide a reliable calculation of the wave functions in the
region of relativistic momenta $p\gtrsim m$.
So, we introduce a cutoff in~\eqref{eq:idefs} in order to avoid possible errors caused by the mentioned
uncertainty. It is obviously that in the quark model
we can calculate a number of nonperturbative parameters~\eqref{eq:omega-defs} only with certain accuracy.
The way of further improvements in the calculation is related in the first place
with more accurate construction of
the bound state wave function at relativistic momenta. In the approach of NRQCD we encounter
analogous difficulties connected with the determination of numerous nonperturbative matrix
elements~\cite{bodwin1}.

Let us note also that the cross section~\eqref{eq:cs} contains the fourth power of the modified
wave function at the origin $\tilde R(0)$ and the strong coupling constant $\alpha_s$.
Thus, small changes of the bound state wave function can lead to substantial changes in
final results for the cross section. The value $\tilde R(0)$ is calculated with sufficiently
high accuracy with the parameters and potential~\eqref{eq:ham0} of the relativistic quark model.
The parameter $|\tilde R(0)|^4$ undergoes essential decrease in comparison with nonrelativistic
value. But other relativistic corrections connected with the functions $F^{(i)}$ ($i=1,2,3$)
in~\eqref{eq:cs} have the opposite effect on the cross section value~\eqref{eq:cs}.
Analytical expression of
nonrelativistic contribution to the cross section which is determined by $F^{(0)}$ coincides
with previous calculations in~\cite{berezhnoy,qiao,ko,li,qiao2002}.
In the evaluation of $\alpha_s$ we set the renormalization
scale to be the transverse mass $\mu=m_T=\sqrt{4m^2+P_T^2}$, which is generally accepted
choice. For the running coupling constant $\alpha_s(\mu)$ we use the LO result with
the initial value $\alpha_s(\mu=M_z)=0.118$.

The basic expression~\eqref{eq:cs-plus-x}
for the calculation of the differential cross section contains
the gluon distribution functions in the proton because the leading contribution comes
from a gluon fusion process. When the energy of colliding beams increases, the initial
parton momentum fraction $x_i$ needed to produce heavy quarkonium decreases. It reaches
the region in $x$ where the number of the gluons becomes much larger than the number
of quarks. The gluon distribution function determines the probability to find a gluon
in the proton with some momentum fraction. There exists a number of the parameterizations
for partonic distribution functions~\cite{CTEQ5L}. We use the gluon PDF from the set CTEQ5L
as in Ref.~\cite{berezhnoy}.
The total numerical value of the cross section obtained from~\eqref{eq:cs} is equal to
\be
\label{eq:total}
\sigma_{total}=9.6~\text{nb}.
\ee
Due to law-x behavior of gluon distribution functions, the main contribution to
the integral cross section~\eqref{eq:total} results from the region $x_{1,2}\sim 10^{-3}$.
To be more precise, this region is determined by the condition:
$7.8\cdot 10^{-7}<x_1x_2< 7.8\cdot 10^{-6}$.
In the frequently used nonrelativistic limit when $\psi(0)=0.21$~GeV$\strut^\frac32$ and
$R(0)=0.74$~GeV$\strut^\frac32$
the total value of the cross section amounts \hbox{$\sigma_{nonrel}=18.3$~nb}
and agrees with the calculation in~\cite{berezhnoy}. In the nonrelativistic limit
of our quark model based on Eqs.~\eqref{eq:ham0} and \eqref{eq:delts} we obtain
slightly greater values $R(0)=0.79$~GeV$\strut^\frac32$ and \hbox{$\sigma_{nonrel}=23.1$~nb}.
To obtain~\eqref{eq:total} the factorization scale in the parton distribution
functions $f_{g/p}(x,\mu)$ is taken equal to the transverse mass too: $\mu=m_T$.

\begin{figure}[t]
\center
\includegraphics{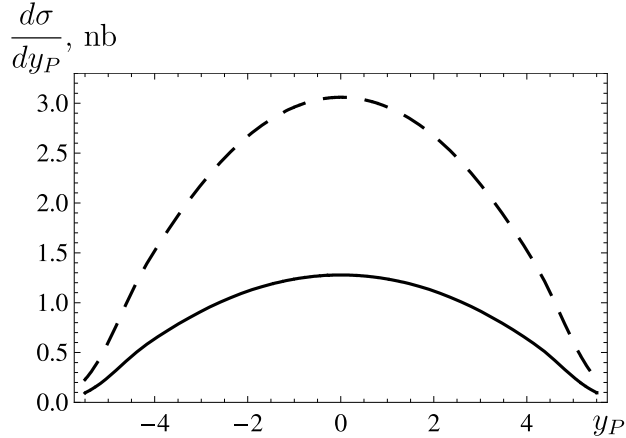}
\caption{The differential cross sections for $pp\to 2J/\psi+X$
at $\sqrt{S}=7$~TeV as functions of rapidity~$y_P$.
Solid and dashed curves represent total and nonrelativistic
results respectively.}
\label{fig4:diags}
\end{figure}

\begin{figure}[t]
\center
\includegraphics{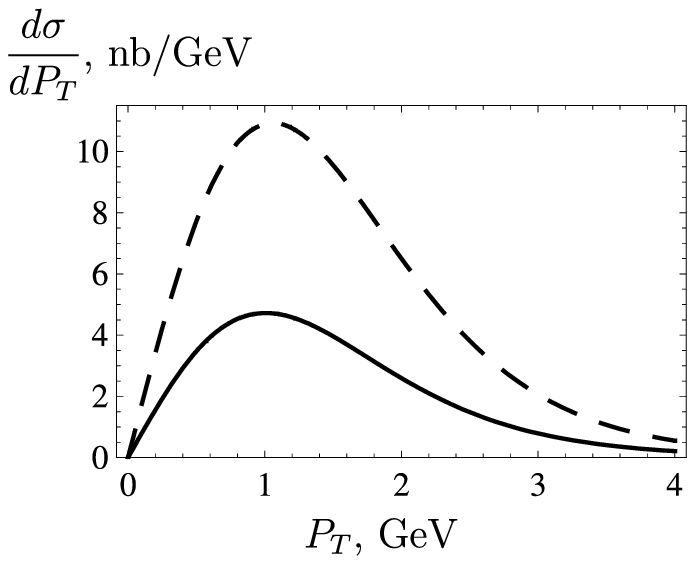}\hfill\includegraphics{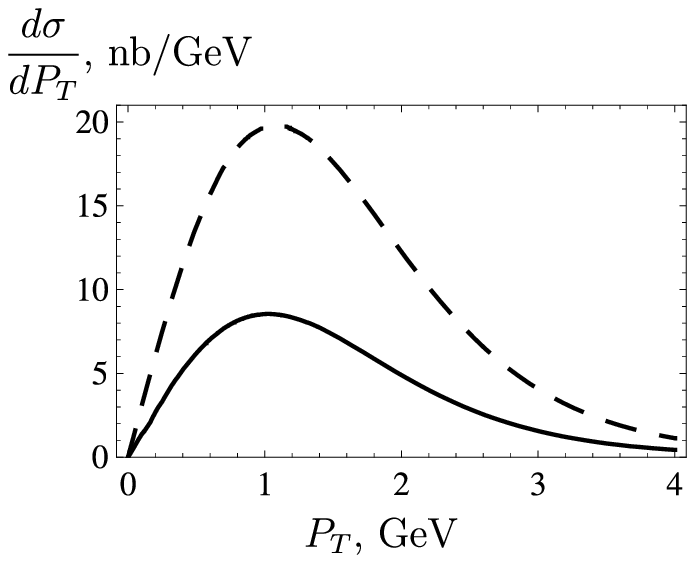}
\caption{The differential cross sections for $pp\to 2J/\psi+X$
at $\sqrt{S}=7$~TeV (left) and $\sqrt{S}=14$~TeV (right)
as functions of transverse momentum $P_T$ of the $J/\psi$ pair integrated over the
rapidity. Solid and dashed curves represent total and nonrelativistic
results respectively.}
\label{fig5:diags}
\end{figure}

To compare the results of our calculation with the measured value of the cross section
in~\cite{LHCb} it is necessary to write the differential cross section in terms of the
rapidity $y_P=\frac{1}{2}\ln\frac{P_0+P_{\parallel}}{P_0-P_{\parallel}}$. The rapidities of outcoming
charmonia with momenta $P$ and $Q$ can be obtained in the form:
\be
\label{eq:raps}
y_{P,Q}=\frac12\ln\frac{x_1}{x_2}\pm\frac12\ln\left[ \frac{s}{M^2-t} -1 \right].
\ee
The differential cross section $d\sigma/dy_P$ for the reaction $pp\to 2J/\psi+X$
is shown in Fig.~\ref{fig4:diags}. It is clear from this plot that relativistic effects strongly
influence on the rapidity distribution of the final charmonium.
In the LHCb experiment~\cite{LHCb} the rapidity lies in the range \hbox{$2<y_{P,Q}<4.5$},
so we should integrate the differential cross section~\eqref{eq:cs-plus-x} over rapidities from
such interval in order to obtain the value corresponding to the experiment at the LHCb detector.
Then we obtain:
\be
\label{eq:rap-cuts}
\sigma^{th}(2< y_{P,Q}< 4.5)=1.6~\text{nb}.
\ee

The result~\eqref{eq:rap-cuts} is significantly smaller than the experimental value of the cross
section~\eqref{eq:lhcb-num}. Different sources of relativistic corrections in~\eqref{eq:cs-plus-x}
are differently directed. But we observe that combined action of all relativistic effects lead
to essential decreasing of the production cross section.
In this work we carry out the investigation only of one important source of corrections
to the nonrelativistic cross section. Decreasing behavior of the cross section $\sigma(pp\to 2J/\psi+X)$
due to the account of relativistic contributions is noticeable clearly in spite of
existing theoretical errors occurred in our calculation.
In our analysis  of the production amplitudes we correctly take into account
relativistic contributions of order $O(v^2)$. Therefore the first basic
theoretical uncertainty of our calculation is related with the
omitted terms of order $O(v^4)$. Since the calculation
of the charmonium mass is sufficiently accurate in our
model (the error is less then $1\%$), we suppose
that the uncertainty in the cross section calculation due to
omitted relativistic corrections of order $O(v^4)$ in the
quark interaction operator (the Breit Hamiltonian) is also very small.
Taking into account that the average value of the heavy quark velocity squared in the
charmonium is $\langle v^2\rangle=0.3$, we expect that relativistic corrections
of order $O(v^4)$ to the cross section~\eqref{eq:rap-cuts} coming from the production
amplitude should not exceed $30\%$ of the obtained
relativistic result. As we mentioned above in the quasipotential approach
we can not find precisely the bound state wave functions in the region of
relativistic momenta $p\ge m$. Using indirect arguments related with the
mass spectrum calculation we estimate in $10\%$ the uncertainty in the wave function
determination. Larger value of the error will lead to the essential
discrepancy between the experiment and theory in the calculation of the charmonium mass
spectrum. Then the corresponding error in the cross section~\eqref{eq:rap-cuts} is not
exceeding $20\%$. We do not consider a part of theoretical error related with
radiative corrections of order $\alpha_s$ because these corrections are omitted in our analysis.
So, our total theoretical error is not exceeding 36$\%$. To obtain this estimate we add the above
mentioned uncertainties in quadrature.

We show in Fig.~\ref{fig5:diags} the distribution over transverse momentum of the
$J/\psi$ mesons integrated over all rapidities at $\sqrt{S}=7$~TeV (left) and $\sqrt{S}=14$~TeV
(right). Previous investigations~\cite{berezhnoy,li,ko}
of the pair charmonium production in $pp$ interaction showed that the color singlet channel
prevails in the differential cross section $d\sigma/dP_T(pp\to 2J/\psi+X)$ at small $P_T$, but
the color octet channel dominates at large $P_T$.
It can be seen in Fig.~\ref{fig5:diags} at $\sqrt{S}=7$~TeV
that the account of relativistic corrections leads to the
ratio of relativistic and nonrelativistic cross sections $\sigma_{rel}/\sigma_{nr}\approx 0.4$ near the
peak. This trend remains unchanged in the region of high transverse momenta. So, the color-octet
contribution retains the dominance at large $P_T$. We investigate also the relative value of
relativistic corrections in the production rate with the growth of the energy $\sqrt{S}$. Our
calculation show that at $\sqrt{S}=14$~TeV (see the right plot in Fig.~\ref{fig5:diags})
the ratio of relativistic and nonrelativistic cross sections
is retained without essential modifications. The cross section increases with the
growth of the energy and reaches the value $\sigma^{th}(2< y_{P,Q}< 4.5)=2.98$~nb at $\sqrt{S}=14$~TeV.
It is appropriate to mention here one result regarding to the study of relativistic effects
in single $J/\psi$ production at hadron collisions~\cite{chao2009}. It was shown
in that paper that relativistic corrections to the color-singlet $J/\psi$ hadroproduction
of order $O(v^2)$ are at a level of about 1$\%$ for sufficiently large $P_T$:
$5\leq P_T\leq 50$~GeV. Our calculation demonstrates that in the region of transverse
momenta $5\leq P_T\leq 50$~GeV the value of relativistic corrections in the cross section
of the pair charmonium production reaches $60\%$.
Relativistic corrections which we study in this work include not only the terms
of order $O(v^2)$ in the production amplitude but also the same order effects in
the long-distance matrix elements. In spite of the difference between~\eqref{eq:lhcb-num}
and~\eqref{eq:rap-cuts},
we consider that at present it is difficult to state that there is the discrepancy between the
theory and experiment in double charmonium production. Indeed, it is known that NLO
in $\alpha_s$ contributions have large value in inclusive single-quarkonium
production at hadron colliders~\cite{brambilla2011,campbell,gong}. The example is found in
the inclusive $J/\psi$ production where the NLO corrections to the color-singlet
contribution increase the total cross section by a factor of about 2 and the
production rate of $J/\psi$ is much increased for larger transverse
momentum $P_T$. Therefore, one
can expect that the NLO corrections to the double charmonium production in proton-proton
interaction can smooth the appeared difference between~\eqref{eq:lhcb-num} and~\eqref{eq:rap-cuts}.
Moreover, as we mentioned
above there exists new mechanism through the double parton scattering which gives
the contribution comparable with the standard nonrelativistic result: $\sigma_{DPS}(pp\to 2J/\psi+X)=
2$~nb~\cite{baranov1}. Accounting for this result and our value of the cross section~\eqref{eq:rap-cuts}
we obtain the summary value $\sigma(pp\to 2J/\psi+X)=3.6$~nb. Then, taking into account
the experimental error, the difference with the LHCb experiment
does not look so significant.

\acknowledgments

The authors are grateful to D. Ebert, R.N. Faustov and V.O. Galkin for useful discussions.
The work is supported partially by the Ministry of Education and
Science of Russian Federation (government order for Samara State U. No. 2.870.2011).

\newpage
\appendix
\section{The coefficients $F^{(i)}$ entering the differential cross section~(\ref{eq:cs})}
\label{app:fis}
\be
\begin{split}
&
F^{(0)}=\frac{16384}{9 M^4 s^6 \left(M^2-t\right)^4 \left(M^2-s-t\right)^4}\Bigl[7776 M^{24}-432 M^{22} (73 s+216 t)+6 M^{20} \times \\&
\negthickspace\left(9085 s^2+60336 s\,t+85536 t^2\right)-16 M^{18} \left(3629 s^3+37686 s^2 t+117855 s\,t^2+106920 t^3\right)+ \\&
2 t^4 (s+t)^4 \left(349 s^4+2304 s^3 t+6192 s^2 t^2+7776 s\,t^3+3888 t^4\right)+4 M^{16} \bigl(11927 s^4+151588 s^3 t+ \\&
745674 s^2 t^2+1470960 s\,t^3+962280 t^4\bigr)-4 M^{14} \bigl(7761 s^5+109608 s^4 t+699467 s^3 t^2+ \\&
2173908 s^2 t^3+3055320 s\,t^4+1539648 t^5\bigr)+2 M^{12} \bigl(6952 s^6+117893 s^5 t+897043 s^4 t^2+ \\&
3741980 s^3 t^3+8278410 s^2 t^4+8872416 s\,t^5+3592512 t^6\bigr)-4 M^2 t^2 (s+t)^2 \bigl(9 s^7+649 s^6 t+ \\&
6460 s^5 t^2+29630 s^4 t^3+74435 s^3 t^4+105156 s^2 t^5+77868 s\,t^6+23328 t^7\bigr)-2 M^{10} \bigl(1899 s^7+ \\&
43398 s^6 t+405618 s^5 t^2+2113568 s^4 t^3+6394090 s^3 t^4+10762584 s^2 t^5+9189936 s\,t^6+ \\&
3079296 t^7\bigr)+M^8 \bigl(587 s^8+19710 s^7 t+244772 s^6 t^2+1603468 s^5 t^3+6229962 s^4 t^4+ \\&
14478304 s^3 t^5+19359816 s^2 t^6+13582080 s\,t^7+3849120 t^8\bigr)-2 M^6 \bigl(20 s^9+1185 s^8 t+ \\&
22153 s^7 t^2+193780 s^6 t^3+965358 s^5 t^4+2928368 s^4 t^5+5431786 s^3 t^6+5949528 s^2 t^7+ \\&
3508920 s\,t^8+855360 t^9\bigr)+M^4 \bigl(s^{10}+76 s^9 t+3756 s^8 t^2+52062 s^7 t^3+353472 s^6 t^4+ \\&
1398834 s^5 t^5+3421754 s^4 t^6+5210968 s^3 t^7+4784622 s^2 t^8+2414880 s\,t^9+513216 t^{10}\bigr)
\Bigr],
\end{split}
\ee

\be
\begin{split}
&
F^{(1)}=-\frac{16384}{27 M^4 s^8 \left(M^2-t\right)^5 \left(M^2-s-t\right)^5}\Bigl[497664 M^{32}-221184 M^{30} (13 s+36 t)+ \\&
5760 M^{28}\bigl(1285 s^2+7680 s\,t+10368 t^2\bigr)-48 M^{26} \bigl(243089 s^3+2289552 s^2 t+6612480 s\,t^2+ \\&
5806080 t^3\bigr)+12 M^{24}\bigl(1090607 s^4+13899232 s^3 t+62988960 s^2 t^2+117411840 s\,t^3+\\&
75479040 t^4\bigr)-8 M^{22}\bigl(1392130 s^5+22255745 s^4 t+136976040 s^3 t^2+399063744 s^2 t^3+\\&
540933120 s\,t^4+271724544 t^5\bigr)+8 t^5 (s+t)^5 \bigl(1867 s^6+18256 s^5 t+77728 s^4 t^2+181152 s^3 t^3+\\&
246096 s^2 t^4+186624 s\,t^5+62208 t^6\bigr)+4 M^{20} \bigl(1800338 s^6+35626541 s^5 t+276305481 s^4 t^2+\\&
1095702384 s^3 t^3+2313080352 s^2 t^4+2435457024 s\,t^5+996323328 t^6\bigr)-M^{18} \bigl(3514643 s^7+\\&
86600280 s^6 t+825013064 s^5 t^2+4139468480 s^4 t^3+11909795760 s^3 t^4+19466599680 s^2 t^5+\\&
16605388800 s\,t^6+5693276160 t^7\bigr)+2 M^{16} \bigl(659715 s^8+19899554 s^7 t+231809132 s^6 t^2+\\&
1426201784 s^5 t^3+5209721940 s^4 t^4+11612866752 s^3 t^5+15333506496 s^2 t^6+\\&
10912112640 s\,t^7+3202467840 t^8\bigr)-M^2 t^3 (s+t)^3 \bigl(664 s^9+73857 s^8 t+969897 s^7 t^2+\\&
6006640 s^6 t^3+21622120 s^5 t^4+49033392 s^4 t^5+71696784 s^3 t^6+65938176 s^2 t^7+\\&
34725888 s\,t^8+7962624 t^9\bigr)-M^{14} \bigl(380999 s^9+13782634 s^8 t+195250554 s^7 t^2+\\&
1459312784 s^6 t^3+6543590240 s^5 t^4+18571278768 s^4 t^5+33449165568 s^3 t^6+\\&
36752348160 s^2 t^7+22298664960 s\,t^8+5693276160 t^9\bigr)+2 M^{12} \bigl(39553 s^{10}+1754120 s^9 t+\\&
30305432 s^8 t^2+275173836 s^7 t^3+1497456520 s^6 t^4+5228976572 s^5 t^5+12018927060 s^4 t^6+
\end{split}
\ee
\bes
\begin{split}
&
18013988160 s^3 t^7+16836887232 s^2 t^8+8856207360 s\,t^9+1992646656 t^{10}\bigr)-4 M^{10}\times\\&
\bigl(2658 s^{11}+152991 s^{10} t+3330028 s^9 t^2+37242554 s^8 t^3+246717845 s^7 t^4+1047413960 s^6 t^5+\\&
2970379604 s^5 t^6+5692479912 s^4 t^7+7256996388 s^3 t^8+5868903744 s^2 t^9+2712213504 s\,t^{10}+\\&
543449088 t^{11}\bigr)+2 M^4 t (s+t) \bigl(8 s^{12}+943 s^{11} t+73235 s^{10} t^2+1243892 s^9 t^3+10501734 s^8 t^4+\\&
53251684 s^7 t^5+176208052 s^6 t^6+396112960 s^5 t^7+612104570 s^4 t^8+641021472 s^3 t^9+\\&
434301120 s^2 t^{10}+171417600 s\,t^{11}+29859840 t^{12}\bigr)+2 M^8 \bigl(267 s^{12}+31089 s^{11} t+948124 s^{10} t^2+\\&
13678946 s^9 t^3+112700707 s^8 t^4+584919678 s^7 t^5+2023301960 s^6 t^6+4798139192 s^5 t^7+\\&
7834428690 s^4 t^8+8641058880 s^3 t^9+6129541440 s^2 t^{10}+2515968000 s\,t^{11}+452874240 t^{12}\bigr)-\\&
M^6 \bigl(8 s^{13}+1772 s^{12} t+139013 s^{11} t^2+3024912 s^{10} t^3+33019131 s^9 t^4+215628282 s^8 t^5+\\&
917347710 s^7 t^6+2664532144 s^6 t^7+5399964368 s^5 t^8+7640313480 s^4 t^9+7391197632 s^3 t^{10}+\\&
4650333696 s^2 t^{11}+1710858240 s\,t^{12}+278691840 t^{13}\bigr)
\Bigr],
\end{split}
\ees
\be
F^{(2)}=-4F^{(0)},
\ee
\be
\begin{split}
&
F^{(3)}=\frac{4096}{81 M^4 s^{10} \left(M^2-t\right)^6 \left(M^2-s-t\right)^6}\Bigl[ 31850496 M^{40}-995328 M^{38} (219 s+640 t)+\\&
27648 M^{36} \bigl(33541 s^2+153432 s\,t+218880 t^2\bigr)-9216 M^{34} \bigl(313031 s^3+1832460 s^2 t+\\&
4240836 s\,t^2+3939840 t^3\bigr)+768 M^{32} \bigl(8244959 s^4+63274344 s^3 t+189539568 s^2 t^2+\\&
295052544 s\,t^3+200931840 t^4\bigr)-192 M^{30} \bigl(52201757 s^5+519109072 s^4 t+2012650416 s^3 t^2+\\&
4109435136 s^2 t^3+4827651840 s\,t^4+2571927552 t^5\bigr)+16 M^{28} \bigl(760108913 s^6+9260081144 s^5 t+\\&
46005056832 s^4 t^2+120457679616 s^3 t^3+188334270720 s^2 t^4+177640667136 s\,t^5+\\&
77157826560 t^6\bigr)-8 M^{26} \bigl(1475952353 s^7+20899343744 s^6 t+127598524184 s^5 t^2+\\&
423537421056 s^4 t^3+844308839808 s^3 t^4+1076247502848 s^2 t^5+846934050816 s\,t^6+\\&
308631306240 t^7\bigr)+16 t^6 (s+t)^6 \bigl(117307 s^8+841072 s^7 t+2960704 s^6 t^2+7010976 s^5 t^3+\\&
12425424 s^4 t^4+16277760 s^3 t^5+14715648 s^2 t^6+7962624 s\,t^7+1990656 t^8\bigr)+4 M^{24}\times\\&
\bigl(2269174169 s^8+37035895336 s^7 t+266823310648 s^6 t^2+1087787332416 s^5 t^3+\\&
2722434981120 s^4 t^4+4414063320576 s^3 t^5+4767985797120 s^2 t^6+3212405194752 s\,t^7+\\&
1003051745280 t^8\bigr)-M^{22} \bigl(5273945643 s^9+102674924584 s^8 t+863595364448 s^7 t^2+\\&
4191387926848 s^6 t^3+12844048277696 s^5 t^4+25933611095040 s^4 t^5+35654616990720 s^3 t^6+\\&
33483324506112 s^2 t^7+19674332061696 s\,t^8+5349609308160 t^9\bigr)+2 M^{20} \bigl(1108300275 s^{10}+\\&
26970163113 s^9 t+270274383346 s^8 t^2+1544514084352 s^7 t^3+5656684280280 s^6 t^4+\\&
13921443451712 s^5 t^5+23690831678976 s^4 t^6+28410755383296 s^3 t^7+23612613792768 s^2 t^8+\\&
12267586805760 s\,t^9+2942285119488 t^{10}\bigr)-M^2 t^4 (s+t)^4 \bigl(82320 s^{11}+10526659 s^{10} t+\\&
122742683 s^9 t^2+663347888 s^8 t^3+2199787352 s^7 t^4+5137832384 s^6 t^5+9050135872 s^5 t^6+\\&
12183171072 s^4 t^7+12114432000 s^3 t^8+8258236416 s^2 t^9+3403026432 s\,t^{10}+637009920 t^{11}\bigr)-\\&
M^{18} \bigl(658318562 s^{11}+20774136911 s^{10} t+256196780079 s^9 t^2+1742958116720 s^8 t^3+\\&
7550194930440 s^7 t^4+22215242834368 s^6 t^5+45850626752448 s^5 t^6+67807235088384 s^4 t^7+
\end{split}
\ee
\bes
\begin{split}
\\&
72355739968512 s^3 t^8+53902515363840 s^2 t^9+25023941369856 s\,t^{10}+5349609308160 t^{11}\bigr)+\\&
4 M^{16} \bigl(34194446 s^{12}+1432273039 s^{11} t+22383564112 s^{10} t^2+185308892030 s^9 t^3+\\&
954476169886 s^8 t^4+3325907679280 s^7 t^5+8190596683920 s^6 t^6+14624279688576 s^5 t^7+\\&
19220073992064 s^4 t^8+18510299039232 s^3 t^9+12475169980416 s^2 t^{10}+5218508685312 s\,t^{11}+\\&
1003051745280 t^{12}\bigr)-M^{14} \bigl(19246515 s^{13}+1103736798 s^{12} t+22478276084 s^{11} t^2+\\&
232094334116 s^{10} t^3+1443234697510 s^9 t^4+5968097322672 s^8 t^5+17386794613376 s^7 t^6+\\&
36925721101440 s^6 t^7+58395873784128 s^5 t^8+69341740385280 s^4 t^9+60881635077120 s^3 t^{10}+\\&
37374891540480 s^2 t^{11}+14182623756288 s\,t^{12}+2469050449920 t^{13}\bigr)+2 M^4 t^2 (s+t)^2\times\\&
\bigl(1024 s^{14}+151570 s^{13} t+12969265 s^{12} t^2+212727487 s^{11} t^3+1644135147 s^{10} t^4+\\&
7651899125 s^9 t^5+24229924530 s^8 t^6+56242255280 s^7 t^7+100001584488 s^6 t^8+\\&
138359684800 s^5 t^9+146996284416 s^4 t^{10}+115115443200 s^3 t^{11}+61947293184 s^2 t^{12}+\\&
20275826688 s\,t^{13}+3025797120 t^{14}\bigr)+2 M^{12} \bigl(824111 s^{14}+70166611 s^{13} t+1952942482 s^{12} t^2+\\&
25962244078 s^{11} t^3+199207601182 s^{10} t^4+988657795894 s^9 t^5+3408779823892 s^8 t^6+\\&
8542312029248 s^7 t^7+16016481697464 s^6 t^8+22847237069376 s^5 t^9+24808543913472 s^4 t^{10}+\\&
19996345055232 s^3 t^{11}+11238532171776 s^2 t^{12}+3890175787008 s\,t^{13}+617262612480 t^{14}\bigr)-\\&
M^6 t (s+t) \bigl(2176 s^{15}+414368 s^{14} t+34529090 s^{13} t^2+736175475 s^{12} t^3+7430265423 s^{11} t^4+\\&
44335186685 s^{10} t^5+175863201255 s^9 t^6+499176650440 s^8 t^7+1064569592864 s^7 t^8+\\&
1756640077312 s^6 t^9+2262689749056 s^5 t^{10}+2242313957376 s^4 t^{11}+1643405663232 s^3 t^{12}+\\&
831421919232 s^2 t^{13}+256940937216 s\,t^{14}+36309565440 t^{15}\bigr)-M^{10} \bigl(48840 s^{15}+\\&
10192723 s^{14} t+435518065 s^{13} t^2+7912226162 s^{12} t^3+77545065352 s^{11} t^4+\\&
471337098042 s^{10} t^5+1940776757542 s^9 t^6+5736471274304 s^8 t^7+12651167529752 s^7 t^8+\\&
21342766010240 s^6 t^9+27846873221696 s^5 t^{10}+27881390733312 s^4 t^{11}+\\&
20726379168768 s^3 t^{12}+10705439195136 s^2 t^{13}+3395959603200 s\,t^{14}+493810089984 t^{15}\bigr)+\\&
4 M^8 \bigl(24 s^{16}+60117 s^{15} t+6480104 s^{14} t^2+187031341 s^{13} t^3+2509040584 s^{12} t^4+\\&
19370940805 s^{11} t^5+97105255536 s^{10} t^6+341150600006 s^9 t^7+884070584925 s^8 t^8+\\&
1749157202600 s^7 t^9+2695987058680 s^6 t^{10}+3249848355136 s^5 t^{11}+3016059149568 s^4 t^{12}+\\&
2073684017664 s^3 t^{13}+987205294080 s^2 t^{14}+288123568128 s\,t^{15}+38578913280 t^{16}\bigr)
\Bigr].
\end{split}
\ees

\end{document}